\documentclass[preprint,12pt]{elsarticle}

\usepackage{amssymb}
\usepackage{amsthm}
\usepackage{amsmath}
\usepackage{bm}

\journal{Annals of Physics}

\begin{document}

\begin{frontmatter}



\title{General Navier-Stokes-like Momentum and Mass-Energy Equations}


\author{Jorge Monreal}
\ead{jmonreal@mail.usf.edu}

\address{Department of Physics, University of South Florida, Tampa, Florida, USA}

\begin{abstract}
A new system of general Navier-Stokes-like equations is proposed to model electromagnetic flow utilizing analogues of hydrodynamic conservation equations.  Such equations are intended to provide a different perspective and, potentially, a better understanding of electromagnetic mass, energy and momentum behavior.  Under such a new framework additional insights into electromagnetism could be gained.  To that end, we propose a system of momentum and mass-energy conservation equations coupled through both momentum density and velocity vectors.   
\end{abstract}


\end{frontmatter}


\section{\label{sec:level1}Introduction}

\subsection{\label{sec:level2}System of Navier-Stokes Equations}

Several groups have applied the Navier-Stokes (NS) equations to Electromagnetic (EM) fields through analogies of EM field flows to hydrodynamic fluid flow.  Most recently, Boriskina and Reinhard made a hydrodynamic analogy utilizing Euler's approximation to the Navier-Stokes equation in order to describe their concept of Vortex Nanogear Transmissions (VNT), which arise from complex electromagnetic interactions in  plasmonic nanostructures \cite{boriskina12}.  In 1998, H. Marmanis published a paper that described hydrodynamic turbulence and made direct analogies between components of the NS equation and Maxwell's equations of electromagnetism\cite{marmanis98}.  Kambe formulated equations of compressible fluids using analogous Maxwell's relation and the Euler approximation to the NS equation\cite{kambe10}.  Lastly, in a recently published paper John B. Pendry, et. al. developed a general hydrodynamic model approach to plasmonics \cite{ciraci13}.
 
In the cases of Kambe and Boriskina, et.al, the groups built their models through analogous Euler-like equations along with relevant mass continuity analogues, respectively shown below.  
 \begin{eqnarray}
\frac{\mathcal{D} \mathbf {v}}{\mathcal{D} t}=-\frac{\nabla p}{\rho},\\
\frac{\mathcal{D} \rho}{\mathcal{D}t} + \rho \nabla \cdot \mathbf{ v}  = 0 
\end{eqnarray}
where $\bf v$ is the velocity vector, $\nabla=\frac{\partial}{\partial x_i} \hat{e}_i$ is the del operator, $p$ is pressure, $\rho$ is fluid density, and $\mathcal{D}/\mathcal{D}t=\partial/\partial t + \mathbf{ v} \cdot \nabla$ is a material derivative operator.  Marminis and others \cite{kauranen12,roux06} utilized the Navier-Stokes equation (3) to build their EM analogues:
 \begin{eqnarray}
\rho\left(\frac{\partial \bf v}{\partial t}+ (\bf v \cdot \nabla) \bf v \right) =-\nabla p + \mu \nabla^2\bf v + \bf f. 
\end{eqnarray}
The terms on the left side of the equation represent the fluid's inertia per volume.  The $\frac{\partial \bf v}{\partial t}$ term represent an unsteady state acceleration, while $ \bf v \cdot \nabla\bf v$ is a non-linear advection term.  On the right hand side, the sum of the pressure gradient, $ \nabla p$, and the viscosity, $\mu \nabla^2\bf v$, represent the divergence of a stress tensor.  Finally, $\bf f$ represents the sum of all other body forces acting on the system.  Equation (3) is the momentum equation that describes fluid flow, while equation (1) is its approximation under zero body forces and inviscid flow, neglecting heat conduction, also termed the Euler approximation. 

As others have done, we, likewise, begin with an analogy of hydrodynamic conservation equations mapped to corresponding electromagnetic conservation equations, assuming non-relativistic flow in an isotropic medium, to finally derive a new system of Navier-Stokes-like equations that model electromagnetic flow.  This new set of equations could potentially be useful in gaining a different perspective and better understanding of electromagnetic mass, energy, momentum behavior.   

\subsection{\label{sec:level2}General Momentum, Mass, Energy Conservation Hydrodynamic Equations}

Equation (3) is not in its most general form to describe fluid momentum.  A more general equation is the Cauchy Momentum equation into which one substitutes in an appropriate stress tensor and constitutive relations relative to the problem at hand.  Such substitution then leads to the NS equation. Making use of the material derivative operator, the Cauchy Momentum Equation is :
 \begin{eqnarray}
\rho \frac{\mathcal{D} \bf v}{\mathcal{D}t} =\nabla \cdot \boldsymbol{\sigma} +  \bf f. 
\end{eqnarray}
where $\nabla \cdot \boldsymbol{\sigma}$ is the divergence of a stress tensor, which can be further broken down into the sum of a pressure tensor, $-\nabla p$, and a deviatoric tensor, $ \nabla \cdot \boldsymbol {\tau}$.  So that, $\nabla \cdot \boldsymbol{ \sigma}=-\nabla p + \nabla \cdot \boldsymbol {\tau}$.  Here we have opted to represent tensors as boldface lower-case Greek letters.\footnote {In component form, the stress tensor can be represented as $\sigma_{ij} = \tau_{ij} +\pi \delta_{ij} $, where $\tau_{ij} $ is the stress deviator tensor that distorts a volume component, while  $\pi \delta_{ij} $ is the volumetric stress tensor that tends to change the volume of a stressed body due to pressure exertion.  Thus, to derive the Navier-Stokes equation from the Cauchy momentum equation a stress tensor of the form: $\sigma_{ij} = -p \delta_{ij} + 2\mu\epsilon_{ij} $ is used, with $\mu\epsilon_{ij}$ representing the viscosity component and $p$ the pressure.}

Given the above, the question then becomes:  What is necessary to generally define a hydrodynamic model obeying Navier-Stokes-type equations.  The answer comes in the form of conservation of momentum, mass and energy.  In terms of the material derivative operator these three are:
\begin{eqnarray}
\textit{Momentum        }&:& \   \rho \frac{\mathcal{D} \bf v}{\mathcal{D}t} - \nabla \cdot \boldsymbol{\sigma} -\mathbf {f} = 0\\
\textit{Mass                }&:& \   \frac{\mathcal{D} \rho}{\mathcal{D}t} + \rho \nabla \cdot \mathbf{ v}  = 0 \\
\textit{Energy  }&:&   \   \frac{\mathcal{D}S}{\mathcal{D}t} - \frac{Q}{T} = 0
\end{eqnarray}
where $Q$ and $T$ are the heat transfer rate and temperature, respectively. The above equations (5-7) plus relative constitutive equations lead to hydrodynamic models for non-relativistic flows within continuum space dynamics.

\section{\label{sec:level1}Electromagnetic ``Flow'' Differential Equations}

\subsection{\label{sec:level2}General Momentum and Mass-Energy Relations}

Comparing analogues of hydrodynamic conservation equations to electromagnetic conservation equations leads to some useful electromagnetic flow relations.  Starting with a comparison to the Cauchy momentum equation, we must first look for a term analogous to the hydrodynamic stress tensor, $\sigma_{ij}$.  While there is still some controversy over correct electromagnetic momentum relations (Abraham vs Minkowski), thus, also controversy over the appropriate form of stress tensor model, a seeming consensus appears to be for using a Minkowski form \cite{ born75, svoboda94}.  In general, the constitutive relations are $\mathbf{D}=\varepsilon_o\mathbf{E} + \mathbf{P}$ and $\mathbf{B}= \mu_o(\mathbf{H}+\mathbf{M})$, where $\mathbf{E}$ and $\mathbf{B}$ are the electric and magnetic fields, respectively; $\mathbf{P}$ and $\mathbf{M}$ are the polarization and magnetization fields, respectively;$\mathbf{D}$ and $\mathbf{H}$ are the displacement and magnetic ``H''  fields, respectively; and $c=1/\sqrt{\varepsilon_{o}\mu_{o}}$ is the speed of light in vacuum.  We have also used $\varepsilon_o$ as permittivity and $\mu_o$ as permeability both of free space.  The Minkowski stress tensor is then of the form 
 \begin{eqnarray}
 \boldsymbol{\tau}=[\mathbf{E}\mathbf{D} + \mathbf{H}\mathbf{B} -\frac{1}{2}\mathbf{I}(\mathbf{E}\cdot \mathbf{D}+\mathbf{H} \cdot \mathbf{B})]
 \end{eqnarray} 
where, $\mathbf{I}$ is the identity matrix.  The third term of equation (8) contains the energy density defined as, $u=\frac{1}{2}(\mathbf{E\cdot D} + \mathbf{H\cdot B})$.  

To derive desired electromagnetic momentum and mass-energy relations without unnecessary mathematical complications, we choose to work in a non-magnetic, negligibly polarizable , isotropic medium.  With the Minkowski stress tensor, EM conservation of linear momentum, derived from forces on a charged particle of arbitrary volume traveling through an EM field, is \cite{novotny06, schwartz72}: 
\begin{eqnarray}
\nabla \cdot \boldsymbol{\tau} =\frac{\partial\mathbf {g}}{\partial t} +\mathbf{f}
\end{eqnarray}     
Here $\mathbf {g}$ is defined as EM field momentum density ( $\mathbf {g} \equiv \mathbf{S}/c^2$ ) and  $\mathbf{S}$ ($=\mathbf{E} \times \mathbf{H}$) is the Poynting vector.   On the left-hand side of (9), $\nabla \cdot \boldsymbol{\tau}$ represents the total momentum flowing through the surface of an arbitrary volume per unit time, while on the right-hand side $\partial\mathbf {g}/\partial t$ represents the rate of change of field momentum density within such volume and $\mathbf{f}$ is the rate of change of mechanical momentum within the volume \cite{schwartz72} imposed by the Lorenz force. Conservation of electromagnetic momentum, equation (9), is analogous to the hydrodynamic conservation of momentum, equation (5), in the following way
\begin{eqnarray}
\frac{\mathcal{D} \bf g}{\mathcal{D}t} -(\mathbf{v}_{em} \cdot\nabla \mathbf{g})- \nabla \cdot \boldsymbol{\tau} + \mathbf{f} = 0 
\end{eqnarray}   
 where the second term on the left side of (10) must be included to balance the material derivative term to obtain equation (9).  We, thus, introduce a new vector term which represents a time-independent momentum density convective acceleration  
\begin{eqnarray}
\mathbf{\mathcal{A}}_g =\mathbf{v}_{em} \cdot\nabla \mathbf{g}  
\end{eqnarray}
where $\mathbf{v}_{em}$ represents the velocity field of the EM field in space-time and is analogous to the hydrodynamic velocity field convective acceleration, $ \bf v \cdot \nabla \bf v $.  Now, $\mathbf{\mathcal{A}}_g$ is a form of force exerted due to flow of momentum density interacting with the velocity field.  In other words, the gradient of the momentum density vector points in the direction of the velocity field vector.  As such, we group the two forces that appear in (10) into one total force so that $\mathbf{\mathcal{F}}=\mathbf{\mathcal{A}}_g-\mathbf{f}$, the first being a time independent convective force, the second being a time dependent rate of change in mechanical momentum.  As shown in equation (15) below, this completes the analogy with the hydrodynamic Cauchy momentum equation (5).

Unfortunately, there is no conservation of mass equation for an EM field.  But, there is a conservation of energy equation for electromagnetic fields and it is given by the following relation \cite{schwartz72}: 
\begin{eqnarray}
\frac{\partial u}{\partial t} + \nabla\cdot\mathbf{S} + \mathbf{j\cdot E} = 0 
\end{eqnarray}
where $u=\varepsilon_{o}\frac{1}{2}(E^2+c^2B^2)$ and $\mathbf{j}$ is charge current, $\mathbf{j}=\rho \mathbf{v}$ with $\rho=qn$ as charge density and $\mathbf{v}$ as charge velocity.  

Utilizing Einstein's non-relativistic mass-energy relationship, $E=mc^2$, we establish a relation between conservation of electromagnetic energy and conservation of mass from hydrodynamic flow by taking equation (6), after expanding the material derivative operator, multiplying it by $c^2$ and setting it equal to equation (12) to get the following relation
\begin{eqnarray}
\frac{\partial u}{\partial t} + \nabla\cdot\mathbf{S} + \mathbf{j\cdot E} = \left\{\frac{\partial \rho}{\partial t} +\nabla\cdot(\rho\mathbf{v})\right\} \times c^2
\end{eqnarray}    
dividing  by $c^2$ and re-arranging terms we get
\begin{eqnarray}
\frac{\partial \rho_{\rm em}}{\partial t}+ \nabla\cdot\mathbf{g}+\frac{1}{c^2}\mathbf{j\cdot E} -\nabla\cdot(\rho_{\rm ns}\mathbf{v}_{em}) =0
\end{eqnarray}    
where $\rho_{\rm em}=\rho_{\rm ed}-\rho_{\rm ns}$ and $\rho_{\rm ed}=u/c^2$, while $\rho_{\rm ns}$ is a material density of the medium and $\mathbf{v}_{em}$ is as before.  Equation (14) has no analogue to hydrodynamic equations.

In summary, we have the following two EM field conservation relations here derived.  These form a system of Navier-Stokes-like equations whereby additional insight could be gained into electromagnetic flows.
\begin{eqnarray}
\textit{Momentum               }&:& \ \frac{\mathcal{D} \bf g}{\mathcal{D}t} - \nabla \cdot \boldsymbol{\tau} -\mathbf{\mathcal{F}} = 0  \\
\textit{Mass-Energy               }&:& \ \frac{\partial \rho_{\rm em}}{\partial t}+ \nabla\cdot\mathbf{g}+\frac{1}{c^2}\mathbf{j\cdot E} -\nabla\cdot(\rho_{\rm ns}\mathbf{v}_{em}) =0 
\end{eqnarray}
Equation (15) is analogous to the Cauchy momentum equation (5).  It is a vector equation that describes the time rate of change of the EM field momentum density, under assumptions made here.  The Mass-Energy equation is a scalar equation that describes the time rate of change of a so-called EM density given by the difference in energy density, $u$, per $c^2$ and a second density obtained from the medium of the EM field. It has no analogue to hydrodynamics.  The electromagnetic momentum and mass-energy equations are coupled through the momentum density vector, $\mathbf{g}$, and the velocity vector, $\mathbf{v}_{em}$, in a similar fashion to hydrodynamic conservation equations.

\subsection{\label{sec:level2} Euler-like equation}
Interestingly, we obtain a Euler-like approximation for equation (15) upon moving $\nabla \cdot \boldsymbol{\tau}$ to the right-hand side. First, let us represent the divergence of Maxwell stress tensor in component notation.  Since it is a second rank tensor we will have the following, 
\begin{eqnarray}
\nabla \cdot \boldsymbol{\tau} &=& \frac{\partial \tau_{ij}}{\partial x_j}\mathbf{e}_i\nonumber \\
&=&\varepsilon_{o}\left\{ E_j\frac{\partial E_i}{\partial x_j} + c^2B_j\frac{\partial B_i}{\partial x_j}-\delta_{ij}\frac{\partial (u/\varepsilon_o)}{\partial x_j}\right\}\mathbf{e}_i
\end{eqnarray}
where the $\mathbf{e}_i$ are basis vectors. Assuming a non-conducting, vacuum medium so that $\mathbf{f}=0$, equation (15) in component notation becomes
\begin{eqnarray}
&&\left(\frac{\partial g_i}{\partial t}+v_j\frac{\partial g_i}{\partial x_j}\mathbf{e}_i\right)\nonumber-v_j\frac{\partial g_i}{\partial x_j}\mathbf{e}_i\\
&&=\varepsilon_{o}\left\{ E_j\frac{\partial E_i}{\partial x_j}\mathbf{e}_i + c^2B_j\frac{\partial B_i}{\partial x_j}\mathbf{e}_i-\frac{\partial (u/\varepsilon_o)}{\partial x_j}\right\}.
\end{eqnarray}
or in vector notation
\begin{eqnarray}
\frac{\partial \mathbf{g}}{\partial t}=\varepsilon_{o}\left\{ (\mathbf{E}\cdot\nabla)\mathbf{E} + c^2(\mathbf{B}\cdot\nabla)\mathbf{B}-\nabla (u/\varepsilon_o)\right\}.
\end{eqnarray}
Since the electromagnetic wave is propagating in a vacuum, from Maxwell's equations $\nabla\cdot\mathbf{E}=\nabla \cdot\mathbf{B}=0$.  After rearranging, equation (19)  becomes
\begin{eqnarray}
\frac{\partial \mathbf{g}}{\partial t}=-\nabla u.
\end{eqnarray}
Now, the energy density term, $u$, on the right hand side of the equation is measured in energy per unit volume, which is also a measure of pressure as a force per unit area.  Through simple dimensional analysis one can ascertain that  
$$
{\rm \frac{Energy}{Volume}}= \frac{F\cdot d}{A \cdot d}=\frac{F}{A} =P
$$
where F is a force, A is unit area and d is distance.  As a consequence, the energy density term can be thought of as a pressure component so that we can let $p_{\rm em}=u$, as a representation of pressure.  With this substitution, the analogy with Euler approximation is evident.  Under non-divergent $\mathbf{v}$ such that $\nabla \cdot \mathbf{v} =0$, Navier-Stokes equation (1) becomes $\frac{\partial \mathbf{v}}{\partial t} = -\frac{\nabla\ p}{\rho}$, which is analogous to (21) below.  
\begin{equation}
\frac{\partial \mathbf{g}}{\partial t} = -\nabla\ p_{\rm em}.
\end{equation}
We have distinguished electromagnetic pressure from NS pressure through the subscript, em.

\section{Conclusion}

We have shown that by starting with a general form of a hydrodynamic momentum conservation equation, Cauchy momentum, we can analogously derive a general form of electromagnetic  momentum conservation.  As aid to construction, we can likewise use the hydrodynamic conservation of mass equation to derive an electromagnetic relation between mass-energy applicable to electromagnetic ``fluid'' flow.  While, by no means, can these generalized EM Navier-Stokes-like equations be applied to any electromagnetic flow at hand, they could be useful if applied carefully to a system.  One must first decide on appropriate flow assumptions regarding steady-state, vorticity, and constitutive relations appropriate to the medium within which EM flow occurs.  In future work, we will apply these EM conservation equations to describe several well-known problems in electromagnetism.

\section*{Acknowledgements}
This work was partially self-supported through funds obtained while teaching at Polk State College and partially supported by University of South Florida Physics Department.  The author thanks Svetlana Boriskina and Zhimin Shi for very useful suggestions during certain stages of this paper.    



\section*{References}
\bibliographystyle{elsarticle-num} 
\bibliography{NSbib2}

\end{document}